\documentclass[11pt]{article}
\usepackage{moriond,epsfig}
\def\Journal#1#2#3#4{{#1} {\bf #2}, #3 (#4)}

\def\PLB{{\em Phys. Lett.}  B}
\def\PRL{\em Phys. Rev. Lett.}
\def\PRD{{\em Phys. Rev.} D}

\def\be{\begin{equation}}
\def\ee{\end{equation}}
\def\bea{\begin{eqnarray}}
\def\eea{\end{eqnarray}}

\def\met{\not\!\!E_T}
\begin{document}
\begin{flushright} {\normalsize FERMILAB-Conf-04/084-E} \end{flushright}
\vspace*{1cm}
\title{SEARCH FOR SINGLE TOP QUARK PRODUCTION AND MEASUREMENTS OF TOP QUARK 
DECAY PROPERTIES AT THE TEVATRON}
\author{A. JUSTE~\footnote{On behalf of the CDF and {D\O} collaborations.}}
\address{Fermi National Accelerator Laboratory, \\
MS 357, P.O. Box 500, Batavia, IL 60510, USA}

\maketitle
\abstracts{We present preliminary results on the search for single top quark production and
measurements of top quark decay properties by the CDF and {D\O} collaborations of the
Fermilab Tevatron collider, using datasets of 108-164 pb$^{-1}$ of proton-antiproton collisions.}

\section{Introduction}
After its discovery by the CDF and {D\O} collaborations~\cite{topdiscovery} in Run I (1992-1996),
the Fermilab Tevatron collider~\footnote{The Tevatron is a proton-antiproton accelerator
colliding beams in Run II at an increased center-of-mass energy of 1.96 TeV (from 1.8 TeV in Run I) and, as
of beginning of 2004, instantaneous luminosities in excess of $6 \times 10^{31}$ cm$^{-2}$s$^{-1}$
(more than three times the Run I record).}
remains the world's only source of top quarks. 
The large top quark mass~\cite{TEVEWWG}, $m_t = 178.0 \pm 4.3$ GeV, raises the
tantalizing possibility that new physics beyond the Standard Model (SM), possibly
related to the breaking of the electroweak symmetry, might lie just above the electroweak
scale. In such case, its effects might be more apparent in the top quark sector than in 
any other sector of the SM.
Therefore, interactions between the top quark and weak gauge bosons become extremely interesting. 
Within the SM, the top quark interaction to a $W$ boson is of the type $V$--$A$
and involves a $b$ quark almost $100\%$ of the time, thus: 
$\Gamma^{\mu}_{Wtb}\propto |V_{tb}|\gamma^{\mu}(1-\gamma_5)$, with the CKM
matrix element $|V_{tb}|\simeq 1$.

Charged current interactions define most of the top quark phenomenology: final state signature determined
by $W$ boson decay modes, large top quark width,
electroweak production of top quarks, W helicity in top quark decays, $t\bar{t}$ spin correlations, etc.
Despite their importance, the present direct experimental 
knowledge~\cite{RunI_Whel,RunI_SpinCorr,RunI_SingleTop} is rather limited due to the
low statistics available in Run I. In Run II, the anticipated large datasets (up to 4 fb$^{-1}$
according to baseline projections) and improved performance of the upgraded CDF and {D\O}
detectors, will allow to perform incisive tests of the SM in the top quark sector.
This paper summarizes the results from the first measurements on the $W$-$t$-$b$ interaction
using Run II datasets of comparable or larger size than those available in Run I.

\section{Electroweak production of top quarks}
The dominant production mechanisms for a single top quark at the Tevatron
involve the exchange of a $W$ boson, either timelike (s-channel, $p\bar{p}\rightarrow t\bar{b}+X$) 
or spacelike (t-channel, $p\bar{p}\rightarrow tq\bar{b}+X$). The estimated cross sections~\cite{nlostop} 
at the Tevatron at $\sqrt{s}=1.96$ TeV are $\sigma_s =0.88^{+0.07}_{-0.06}$ pb and 
$\sigma_t =1.98^{+0.23}_{-0.18}$ pb, respectively for the s- and t-channels.
With a rate proportional to $|V_{tb}|^2$, single top quark production will allow to measure directly
the CKM matrix element $|V_{tb}|$, thus providing direct experimental verification of the hypothesis of
unitarity for the CKM matrix.

Despite the expected large rate ($\sim 40\%$ of $\sigma_{t\bar{t}}$), the electroweak production of top
quarks has not been observed yet. Existing Run I limits~\cite{RunI_SingleTop} at $95\%$ Confidence Level (CL)
are $\sigma_s < 18(17)$ pb, $\sigma_t < 13(22)$ pb and $\sigma_{s+t} < 14$ pb, as published by the CDF({D\O}) 
collaborations. The experimental signature consists on 
one isolated high $p_T$ isolated electron 
or muon, high transverse missing energy and two or more jets, at least
one of them being a $b$ jet. The dominant background processes are,
in order of importance, W+jets, $t\bar{t}$ and multijets (where a jet is
misidentified as an isolated lepton).

\subsection{Search for single top quark production at CDF}

The CDF Collaboration has performed a search for single top quark production
in the lepton+jets channel using a sample of 162 pb$^{-1}$ collected in Run II. 

A combined (both s- and t-channels) search and a dedicated t-channel 
search have been carried out. Both analyses start with a preselection requiring
one isolated electron or muon with $p_T>20$ GeV and $|\eta_{det}|<1.1$,  $\met >20$ GeV,
exactly 2 jets with $p_T>15$ GeV and $|\eta_{det}|<2.8$, at least one of which is
required to be $b$-tagged using a secondary vertex algorithm, followed by a topological
selection requiring 140 GeV$\leq M_{l\nu b} \leq$ 210 GeV. In addition, in case of the t-channel
search, the leading jet is required to have $p_T>30$ GeV.
Good agreement is found between observation and expectation,
with a total of 28 (25) events observed versus $27.8 \pm 4.3$ ($24.3 \pm 3.5$) expected in
the combined (t-channel) search.

\begin{figure}
\begin{center}
\begin{tabular}{cc}
\includegraphics[width=6cm]{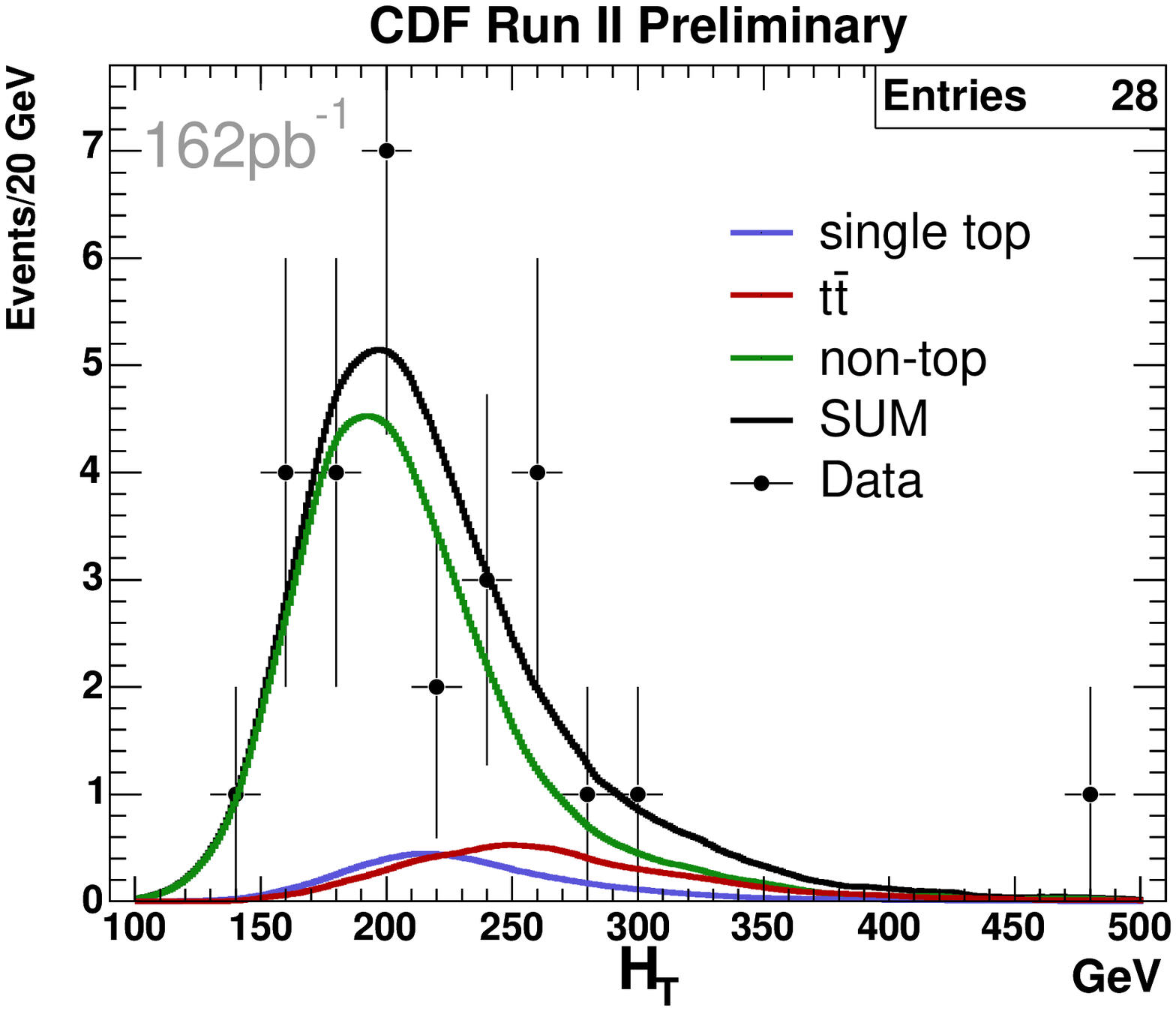} &
\includegraphics[width=6cm]{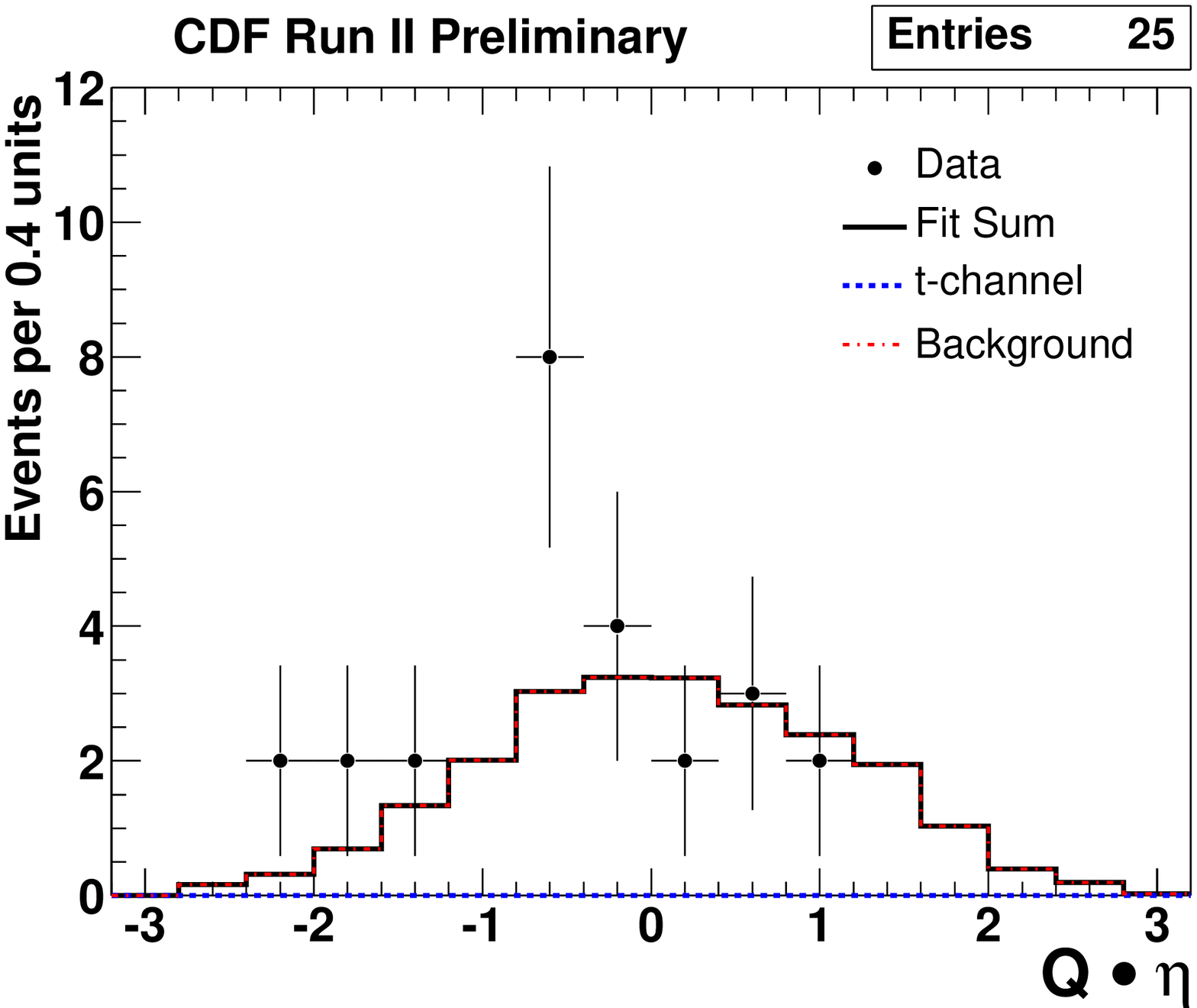} 
\end{tabular}
\end{center}
\caption{$H_T$ (left) and $Q\times\eta$ (right) distributions in data, compared
to the weighted sum of templates corres\-pon\-ding to various contributions as given
by the likelihood fit.
\label{fig:cdf_stop}}
\end{figure}

In order to increase the statistical sensitivity to the signal, a maximum likelihood fit to 
a discriminant variable in data is performed, using a sum of templates determined from Monte Carlo.
In case of the combined search, the variable chosen is $H_T$, defined as the
sum of the lepton $p_T$, $\met$ and jet $p_T$, which shows a similar distribution for both
s- and t-channel processes. In case of the t-channel search, the fit is performed to 
$Q\times\eta$, with $Q$ being the lepton charge and $\eta$ the pseudorapidity of the untagged jet,
which shows a distinct asymmetry for single top t-channel production.
In the fit, the background is allowed to float but is constrained to the expectation.
The actual fit parameters are the deviations with respect to the SM cross-sections, i.e.
$\beta_i = \sigma_i/\sigma^{SM}_i$, with $i$=single top, $t\bar{t}$ and non-top.
The fitted signal content in data (see Fig.~\ref{fig:cdf_stop}) 
are found to be compatible with zero in both searches: 
$\hat{\beta}_{s+t}=0.64 \pm 1.55$ (combined) and $\hat{\beta}_{t}=0.00 \pm 1.39$ 
(t-channel). An upper limit on the single top cross-section is determined from a Bayesian approach
using the likelihood and a flat prior on $\beta$. Systematic uncertainties have been taken into account via a 
convolution procedure. The single top cross-section limits at 95$\%$ CL observed in data are
$\sigma_{s+t}<13.7$ pb and $\sigma_t<8.5$ pb, in good agreement with the expected 
limits of $\sigma_{s+t}<14.1$ pb and $\sigma_t<11.3$ pb, respectively.
\subsection{Search for single top quark production at {D\O}}
The {D\O} Collaboration has also performed a search for single top quark production
in the lepton+jets channel using a sample of 164 pb$^{-1}$ collected in Run II. 

In order to maximize the acceptance for signal, a rather loose event preselection is
applied: one electron (muon) with 
$p_T>15$ GeV and $|\eta_{det}|<1.1$($|\eta|<2.0$),  $\met >15$ GeV, between 2 and 4 jets
with $p_T>15$ GeV and $|\eta_{det}|<3.4$, with the leading jet having 
$p_T>25$ GeV and $|\eta_{det}|<2.5$. In addition, at least one jet is required to be
$b$-tagged, where two classes of algorithms, secondary vertex (SVT) and soft-muon (SLT)
tagging, are considered. Therefore, four orthogonal analysis channels are defined:
$e$+jets/SVT, $\mu$+jets/SVT, $e$+jets/SLT and $\mu$+jets/SLT.
In case of the $e$+jets channel, a jet lifetime probability algorithm has also been used
as a cross-check. Good agreement is found between 
expectation and observation for each separate analysis channel, both in terms of normalization and
shape of kinematic distributions.

This analysis has not applied a final selection yet
and therefore no observed limit in the data has been reported at this conference. 
The expected limits at the preselection level, from the combination of all four orthogonal
analysis channels, are $\sigma_{s+t}<15.8$ pb, $\sigma_s<13.8$ pb and $\sigma_t<19.8$ pb.
The above limits, which are already better than in {D\O} Run I~\cite{RunI_SingleTop}, 
have been obtained using the Modified Frequentist method~\cite{cls} and 
include systematic uncertainties. The expected limits without systematic uncertainties are
about a factor of two smaller. The main systematic uncertainties arise from the determination of
$b$ tagging efficiency in data, jet energy scale, trigger efficiency modeling and assumptions
on the flavor composition of the $W$+jets background.
Significant improvements to the sensitivity are expected from a reduction in the
systematic uncertainties as well as the addition of a final election to the analysis.

\section{Measurement of B($t \rightarrow Wb$)/B($t \rightarrow Wq$) at CDF}
The CDF Collaboration has performed a measurement of the ratio
$b$=B($t \rightarrow Wb$)/B($t \rightarrow Wq$) using a sample of 108 pb$^{-1}$ collected in Run II.
This measurement allows to test the SM prediction of
B($t \rightarrow Wb$)$\simeq 1$. The analysis closely follows the approach used in Run I~\cite{cdf_br},
based on the examination of the $b$-tagging rates in the lepton+jets channel.

The preselected sample is split into four separate subsamples: $=3$-jet and $\geq 4$-jet, each of them
with single and double $b$-tags.
In each of the subsamples, the average event tagging probability for $t\bar{t}$ can be expressed as
a function of $b\epsilon$, where $\epsilon$ represents the single-$b$ tagging efficiency.
Additional effects such as the limited acceptance for $b$-tagging, non-$b$ quark tag
rates, non-$t\bar{t}$ tagged backgrounds, etc, are taken into account.
A likelihood fit is performed to the four subsamples simultaneously to determine the value of
$b\epsilon$ most consistent with the observation. The total number of $t\bar{t}$ events is also
fitted. 
The most likely value for $b\epsilon$ found is $0.25^{+0.22}_{-0.18}$. Using
a single-$b$ tag efficiency of $\epsilon=0.45\pm 0.05$, as measured in calibration samples, one
can infer $b=0.54^{+0.49}_{-0.39}$, consistent with the SM value of 1. This analysis can
also be used to set a lower limit of $b>0.15$ at $95\%$ CL. The same analysis, extrapolated to
500 pb$^{-1}$, is expected to yield a lower limit in $b$ of about 0.7 at $95\%$ CL.

\section{Measurement of the W helicity in top quark decays at {D\O}}
The chiral structure of the $W$-$t$-$b$ vertex in the SM results in $\simeq 70\%(30\%)$ of $W$ bosons
in top quark decays being longitudinally (left-hand) polarized.

The {D\O} Collaboration has performed a measurement of the longitudinal helicity fraction, $F_0$, 
using a sample of 125 pb$^{-1}$ collected in Run I.
This analysis starts from a preselected lepton+jets sample~\cite{d0_mt} and makes additional requirements:
exactly four jets and a cut on the background probability (see below) of less than $10^{-11}$,
resulting in a total of 22 events, with an expected signal-to-background of 1:1. 
The extraction of $F_0$ is based on a likelihood fit to the full event topology,
where the event probability is defined as the normalized 12-fold differential cross-section, computed from
the signal ($t\bar{t}$) and background ($W+4j$) leading order matrix elements, and taking into account
the per-event resolution effects. Optimal treatment of the combinatorial background is achieved by summing
over all 12 possible jet-parton assignments. This method had been successfully applied to the top quark mass
extraction~\cite{d0_mt}, yielding the world's single most precise measurement.
The result obtained is $F_0=0.56\pm 0.31\:{\rm (stat+m_t)}\:\pm 0.07\:{\rm (syst)}$, in good agreement
with the SM expectation $(F_0)_{SM}\simeq 0.7$.
\section{Summary}
The Tevatron collider experiments, CDF and {D\O}, are in a unique position to provide incisive tests of the
SM as well as to search for new physics in the top quark sector. Their extended capabilities as a result of the
detector upgrades, together with the $\geq 20$-fold increase in integrated luminosity with respect to
Run I expected by the end of Run II, open an extremely rich program of measurements of top
quark properties. Here we have shown preliminary results on the search for electroweak production of top
quarks and measurements of top quark decay properties such as 
B($t \rightarrow Wb$)/B($t \rightarrow Wq$) and $F_0$ using datasets of 108-164 pb$^{-1}$.

\end{document}